\documentclass{llncs}
\usepackage{makeidx}  
\usepackage{graphicx}
\usepackage{subfigure}
\usepackage{amssymb,amsmath,bm}
\usepackage{mathrsfs}
\usepackage{booktabs}
\usepackage{multirow}
\usepackage{float}
\usepackage{textcomp}
\usepackage{color}
\begin{document}
	\title{Agent with Warm Start and Active Termination for Plane Localization in 3D Ultrasound}
	
	\author{Haoran Dou$^{1,2\dagger}$ \and Xin Yang$^{3\dagger}$ \and Jikuan Qian$^{1,2}$ \and Wufeng Xue$^{1,2}$ \and Hao Qin$^{1,2}$ \and \\Xu Wang$^{1,2}$ \and Lequan Yu$^{3}$ \and Shujun Wang$^{3}$ \and Yi Xiong$^4$ \and \\Pheng-Ann Heng$^{3}$ \and Dong Ni$^{1,2*}$}
	
	
	\institute{National-Regional Key Technology Engineering Laboratory for Medical Ultrasound, Guangdong Key Laboratory for Biomedical Measurements and Ultrasound Imaging, School of Biomedical Engineering, Health Science Center, Shenzhen University, Shenzhen, China\and
		Medical UltraSound Image Computing (MUSIC) Lab\and
		Department of Computer Science and Engineering, The Chinese University of \\Hong Kong, Hong Kong, China\and
		Department of Ultrasound, Luohu People’s Hospital, Shenzhen, China}
	
	\maketitle
	\let\thefootnote\relax\footnotetext{$\dagger$ authors contribute equally. *Corresponding author: nidong@szu.edu.cn}

\begin{abstract}
	Standard plane localization is crucial for ultrasound (US) diagnosis. In prenatal US, dozens of standard planes are manually acquired with a 2D probe. It is time-consuming and operator-dependent. In comparison, 3D US containing multiple standard planes in one shot has the inherent advantages of less user-dependency and more efficiency. However, manual plane localization in US volume is challenging due to the huge search space and large fetal posture variation. In this study, we propose a novel reinforcement learning (RL) framework to automatically localize fetal brain standard planes in 3D US. Our contribution is two-fold. First, we equip the RL framework with a landmark-aware alignment module to provide warm start and strong spatial bounds for the agent actions, thus ensuring its effectiveness. Second, instead of passively and empirically terminating the agent inference, we propose a recurrent neural network based strategy for active termination of the agent's interaction procedure. This improves both the accuracy and efficiency of the localization system. Extensively validated on our in-house large dataset, our approach achieves the accuracy of $3.4mm/9.6^{\circ}$ and $2.7mm/9.1^{\circ}$ for the transcerebellar and transthalamic plane localization, respectively. Our proposed RL framework is general and has the potential to improve the efficiency and standardization of US scanning.
\end{abstract}

\begin{figure}[h]
	\centering
	\subfigure[ ]{\includegraphics[width=0.28\textwidth]{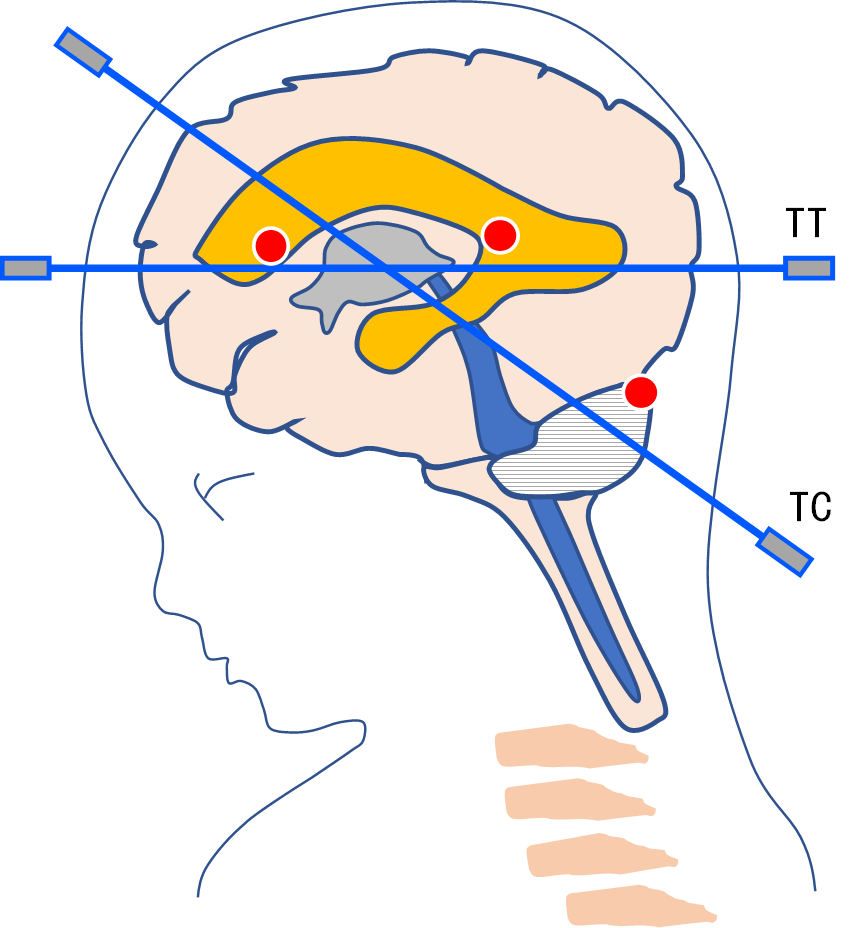}}
	\hspace{.1in}
	\subfigure[ ]{\includegraphics[width=0.31\textwidth]{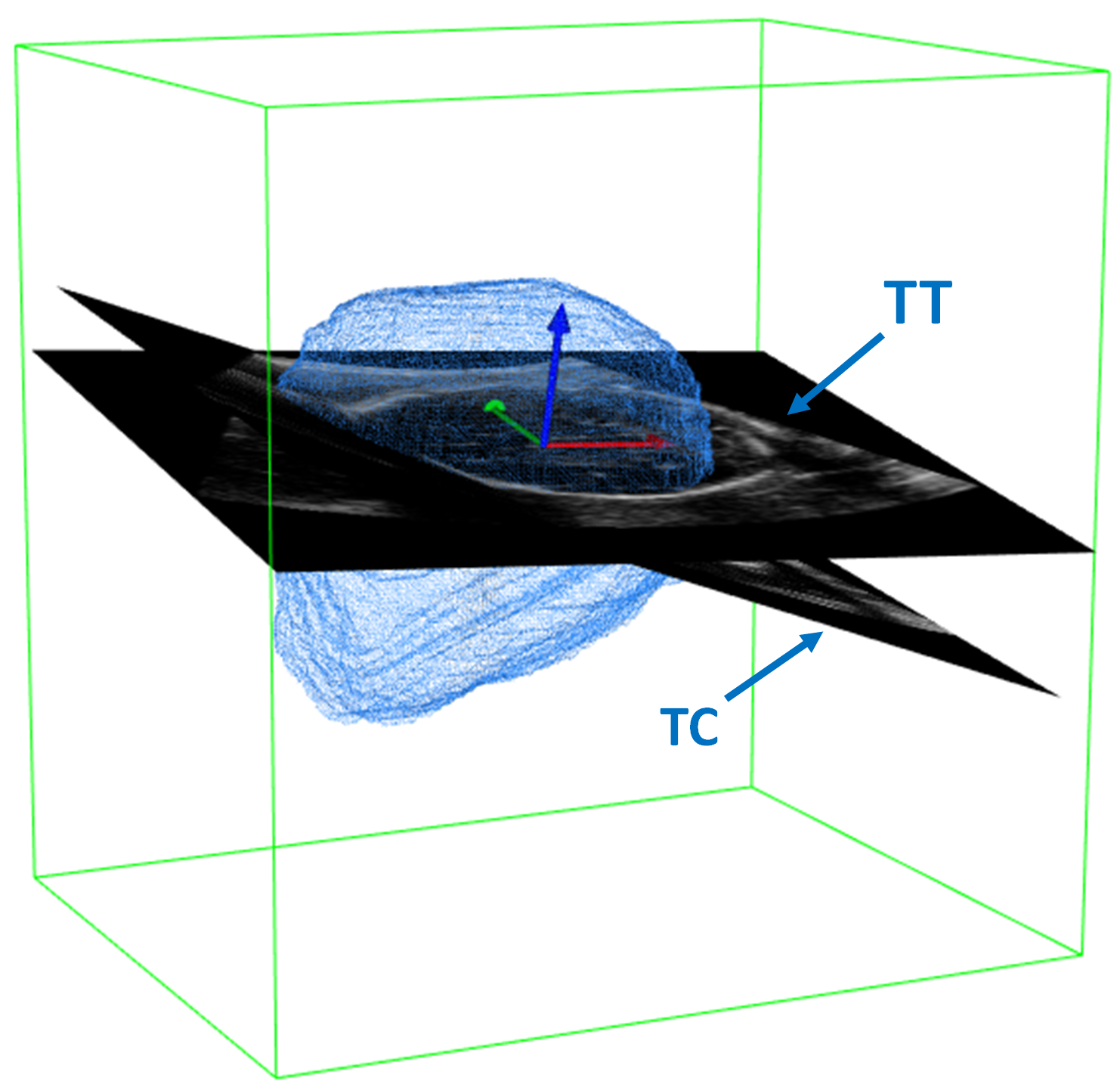}}
	\hspace{.1in}
	\subfigure[ ]{\includegraphics[width=0.325\textwidth]{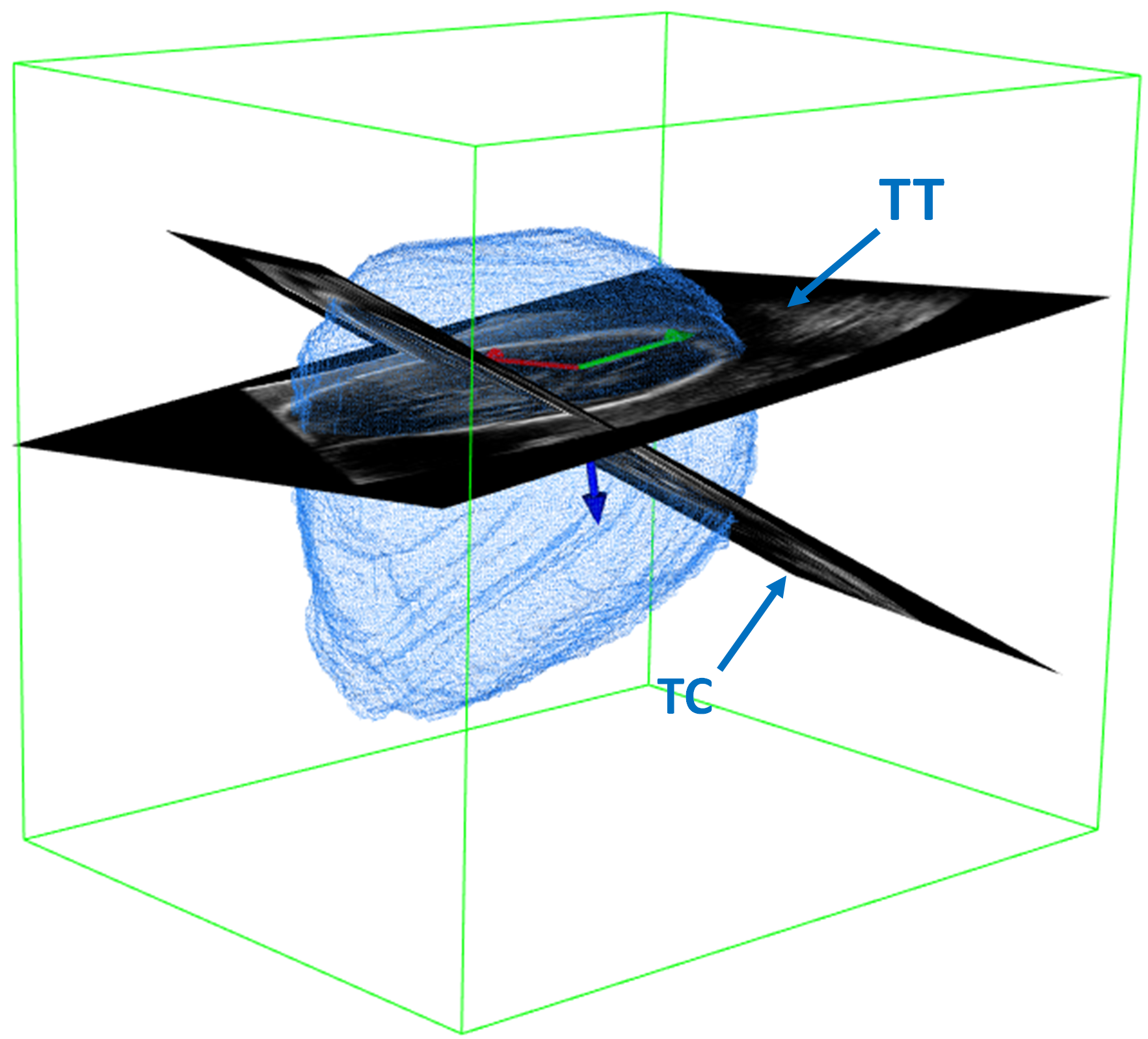}}
	\caption{Fetal brain planes in 3D US. (a) Blue lines show transthalamic (TT) and transcerebellar (TC) plane positions. Red dots (left to right) show three landmarks: genu of corpus callosum, splenium of corpus callosum, cerebellar vermis. (b)(c) Example planes from two volumes illustrate the huge search space and large fetal posture variation.}
	\label{fig:challenge_show}
	\vspace{-0.5cm}		
\end{figure}

\section{Introduction}
Acquisition of standard planes containing key anatomical structures is crucial for ultrasound (US) diagnosis. In prenatal US, typically dozens of standard planes are manually acquired for subsequent biometric measurements and diagnosis with a 2D US probe, such as the transthalamic (TT) and transcerebellar (TC) views for fetal brain assessment (Fig.~\ref{fig:challenge_show}). This process is very time-consuming and highly operator-dependent. In comparison, 3D US can contain multiple standard planes in just a single shot and has the inherent advantages of less user-dependency and more efficiency~\cite{namburete2014diagnostic}. However, it is very challenging to manually localize standard planes in the volume due to the huge search space, the large fetal posture variability and the low image quality, as shown in Fig.~\ref{fig:challenge_show}. Therefore automatic localization of standard planes in 3D US is highly expected to improve diagnostic efficiency and decrease operator-dependency.

In recent years, some research on standard plane localization in 3D US has been conducted accordingly. Ryou et al. proposed a three-step learning method to sequentially localize the fetus, the fetal parts and detect biometry planes by classification~\cite{ryou2016automated}. This method narrowed the search space in the localized structures and the axial direction. Regression methods were also employed to localize cardiac planes by Random Forests~\cite{chykeyuk2013class} and the fetal abdominal plane by deep networks~\cite{schmidt2019offset}. However, these methods tend to fail when acoustic shadow and occlusion spread in US during late pregnancy. Lorenz et al. proposed to extract the abdomen plane by detecting anatomical landmarks and aligning them to a fetal organ model~\cite{lorenz2018automated}. The system achieved accuracy of $5.8mm/15.9^{\circ}$ for plane localization. Although effective by using prior anatomical knowledge, the method's performance is limited by landmark detection accuracy and testing case-model difference. More recently, Li et al. proposed an iterative deep network to localize fetal brain planes in 3D US~\cite{li2018standard}. They further customized a reinforcement learning (RL) agent for view planning in MR volumes~\cite{alansary2018automatic}. RL is promising for standard plane localization in 3D US due to its ability of mimicking experts' operation and exploring inter-plane dependency by the agent-environment interaction. However, RL may suffer from its random initialization and empirical termination when its environment, such as the US volume, has strong noise, artifacts and large appearance variations. \par

In this paper, we propose a novel RL framework to localize fetal brain standard planes in prenatal US volumes. We believe we are the first to employ RL-based techniques for this problem. Our contribution is two-fold. First, we equip the RL framework with a landmark-aware alignment module for warm start to ensure its effectiveness. We employ deep networks to detect anatomical landmarks in the US volume and register them to a plane-specific atlas. The plane configuration of the atlas therefore provides strong spatial bounds for RL agent actions. Second, instead of passively and empirically terminating the agent inference, we propose a recurrent neural network (RNN) based strategy for active termination of the agent's interaction procedure. The RNN-based strategy can find the optimal termination point adaptively, so it improves the accuracy and efficiency of the localization system at the same time. \par
	
\begin{figure}[h]
	\centering
	\includegraphics[width=0.9\linewidth]{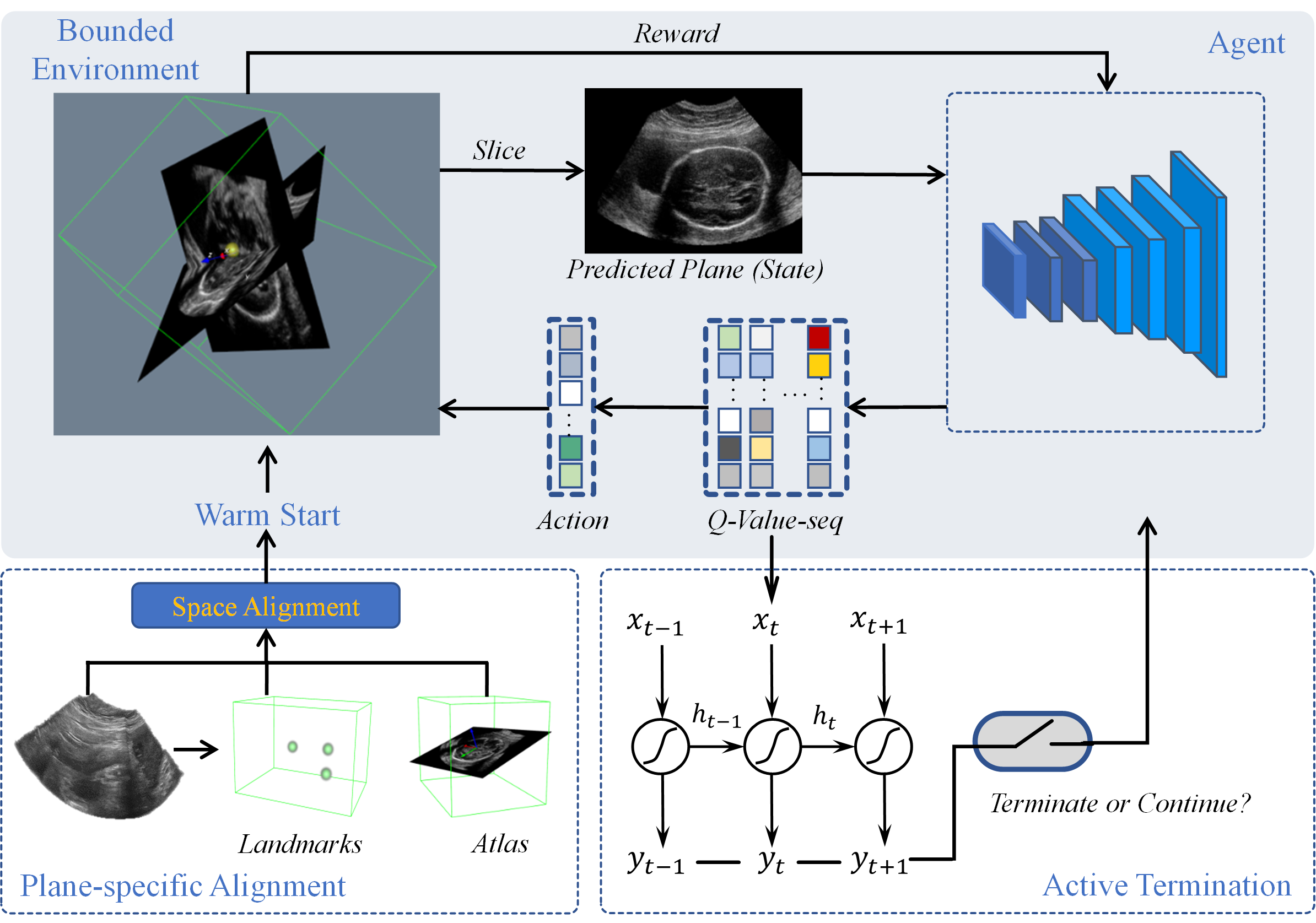}
	\caption{Schematic view of our proposed framework.}
	\label{fig:framework}
	\vspace{-0.6cm}
\end{figure}
			
\section{Methodology}
	
Fig. \ref{fig:framework} is the schematic view of our proposed framework. We propose to localize fetal brain standard planes in US volumes with a RL framework, which can progressively interact with the volumes and modify the search trajectory towards the final target plane. Specifically, we equipped the RL framework with 1) a landmark-aware alignment module for warm start, to ensure its effectiveness, and also 2) a recurrent neural network based strategy for active termination of the interaction procedure, to improve its accuracy and efficiency.
	
\subsection{Deep Reinforcement Learning Framework for Plane Localization}
	
The task of plane localization in US volumes can be well modeled under the RL framework, where an agent, in its current state $s$, interacts with the environments $\mathcal{E}$ by making successive actions $a\in \mathcal{A}$ that maximize the expectation of reward. Let a plane in Cartesian coordinate system be represented as $cos(\alpha)x+cos(\beta)y+cos(\phi)z+d=0$, where $\vec{n}=(cos(\alpha), cos(\beta), cos(\phi))$ is the normal, $d$ is the distance from the plane to the volume center origin. The system will obtain the optimal plane parameters as the agent interacts with the environment.
	
Similar to \cite{alansary2018automatic}, we define the action space as 8 actions, $\{\pm a_{\alpha},\pm a_{\beta},\pm a_{\phi},\pm a_{d}\}$. After an action is made by the agent, the plane parameters are accordingly updated as $\alpha_t=\alpha_{t-1}+a_{\alpha}, \beta_t=\beta_{t-1}+a_{\beta}, \phi_t=\phi_{t-1}+a_{\phi}, d_t=d_{t-1}+a_{d}$. Each valid action gets its scalar reward $r$ following the rule $r=sgn(D(P_{t-1},P_{g})-D(P_{t},P_{g}))$, where $D$ calculates the Euclidean distance from the predicted plane $P_{t}$ to the ground truth $P_{g}$. $r\in\{+1,0,-1\}$ indicates whether the agent is moving towards the preferred target.
	
With the reward signal, the agent then maximizes both the current and future rewards to obtain the action-selection policy. Following the Q-learning, Deep Q-Network \cite{mnih2015human} (DQN) can learn a state-action value function, $Q(s,a)$, via deep networks to serve as the action-selection policy. To improve the robustness of DQN against the noisy environment $\mathcal{E}$ in 3D US, we finally choose the Double DQN (DDQN) \cite{ddqn2016} as our deep agent for plane localization. The loss function for our DDQN is defined as:
\begin{align}
\label{eq:DDQN_loss}
\mathcal{L}_{DDQN}(w) = E_{s,r,a,\hat{s}\sim M}\big[\big(r+\gamma\max \limits_{\hat{a}}Q(\hat{s},Q(\hat{s},a;w);\tilde{w})-Q(s,a;w)\big)^{2}\big],
\end{align}
where $\gamma=0.9$ is a discount factor to weight future rewards, $\hat{s}$ and $\hat{a}$ are the state and action in the next step. $M$ is the experience replay memory to avoid frequent data sampling. $w$ and $\tilde{w}$ are the current and target network parameters. Specifically, we select an ImageNet pre-trained \textit{VGG-13} as our current and target networks. Three recently predicted planes serve as the network input. \par

\begin{figure}[h]
	\centering
	\includegraphics[width=1.0\textwidth]{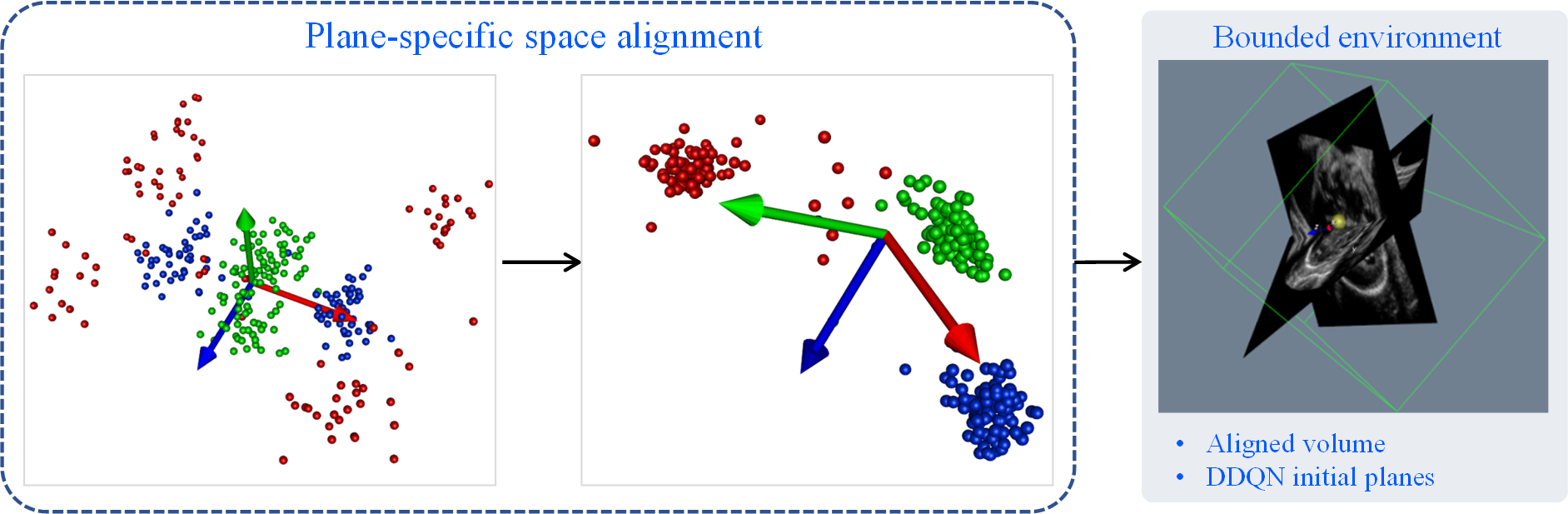}
	\caption{Landmarks of 100 US volumes (left) aligned to a place-specific atlas space (middle) provides strong spatial bounds for RL agent actions (right). Red, green and blue dots indicate landmarks shown in Fig. 1(a).}
	\label{fig:landmark_align}
	\vspace{-0.5cm}
\end{figure}

\subsection{Landmark-aware Plane Alignment for Warm Start}
To ensure an effective interaction of the RL agent with the noisy 3D US environment, we propose a landmark-aware plane alignment module to leverage anatomical prior and provide a warm start for the agent. Specifically, we first detect three landmarks of fetal brain, i.e, the genu of corpus callosum, splenium of corpus callosum and cerebellar vermis, as shown in Fig. \ref{fig:challenge_show}(a), with a customized 3D U-net\cite{ronneberger2015u}. Then these landmarks are used to align the testing volume with the atlas, which contains both the reference landmarks and standard plane parameters. Different from \cite{namburete2014diagnostic,lorenz2018automated} which apply a common anatomical model to all kinds of standard planes, we propose to select specific atlas for each plane to improve the localization accuracy. Finally, standard planes of atlas are mapped to testing volumes and serve as a warm start for our RL agent. Atlas selection for a type of standard plane $P$ is formulated as following,
{\setlength\abovedisplayskip{1pt plus 3pt minus 2pt}
	\setlength\belowdisplayskip{1pt plus 3pt minus 3pt}
\begin{align}
\label{eq:atlas_selection}
\mathcal{X}_{P}^{atl} = \min \limits_{i}\sum_{i}^{N}\sum_{j}^{N}\Big(\Theta(\mathrm{T_{j}^{i}}\times \vec{n}_{P}^{j}, \vec{n}_{P}^{i}) + \parallel d_{P}^{j}-d_{P}^{i}\parallel_{1}\Big), i\neq j.
\end{align}
where $i,j$ are volume index. $\vec{n}_{P}^{i}$ is the normal of $P$ in volume $i$, $\Theta$ calculates the angle between normals, $d_{P}^{i}$ is the distance from plane $P$ in volume $i$ to origin. $\mathrm{T_{j}^{i}}$ is the transformation matrix from volume $j$ to $i$, which is determined by the landmark annotation based rigid registration. Fig.\ref{fig:landmark_align} shows the effect of our landmark alignment for 100 volumes. The accurate alignment guarantees the effectiveness of the initial point for RL agent and therefore leads to fast and improved plane localization. \par	
\begin{figure}[b]
\centering
\includegraphics[width=1.0\linewidth]{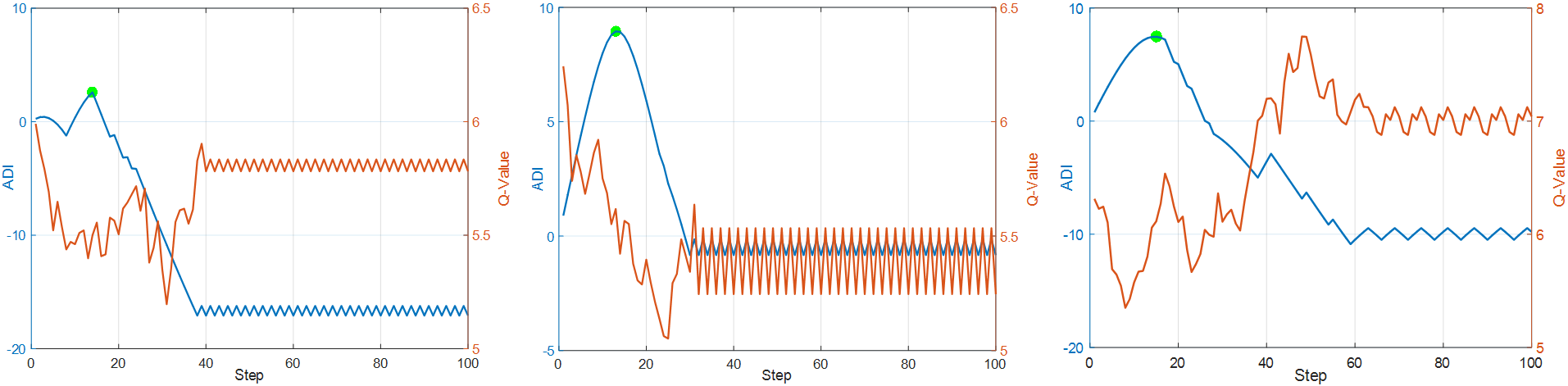}
\caption{Mean Q-value of 8 action candidates (yellow) and ADI (blue) on training dataset. Green point denotes the optimal termination step with maximum ADI.}
\label{fig:q_yes_observation}
\end{figure}	
	
\subsection{Recurrent Neural Network based Active Termination}\label{section:active_terminate}
To ensure an efficient interaction of the RL agent, we propose a RNN-based active termination (AT) module to tell the agent when to stop. Usually, there is no well-defined criteria to terminate the iterative inference of RL learning. Under- and over-estimation of the termination state often degrade the final localization. Existing work makes use of a predefined maximum step, a lower Q-value  \cite{alansary2018automatic} or oscillation of Q-value \cite{ghesu2019multi} as an indicator of termination. While the first one wastes a lot of computation resource if it's set to a large number, the latter two do not necessarily lead to the optimal results. As shown in Fig. \ref{fig:q_yes_observation}, the optimal termination step with highest angle and distance improvement (ADI) is neither the maximum step nor the step with the lowest Q-value. This motivates us to propose a novel strategy to actively learn the optimal step. Specifically, considering the sequential characteristics of the iterative inference, as shown in Fig. \ref{fig:framework}, we formulate the mapping between the Q-value sequence and optimal step with recurrent neural networks. \par
	
The Q-values of 8 action candidates at each state serve as an input of our RNN, which then learns to output the optimal termination step with highest ADI, i.e., most significant angle and distance improvement. We train the RNN model with the inference results of the agent on our training volumes. During testing, our method terminated the iteration action of the agent according to the RNN output and get the final plane parameters. With this active termination mechanism, our agent can make efficient inference without excessive iterations. \par
		
\section{Experimental Results} \label{section:experiment}
\subsubsection{Materials and Implementation Details.} We validate our solution on the task of localizing two standard planes (TT and TC) of fetal brain in US volumes. We built a dataset of 430 prenatal US volumes acquired from 430 healthy pregnant women volunteers. Approved by local Institutional Review Board, all volumes were anonymized and obtained by experts using a Mindray DC-9 US system with an integrated 3D probe. Free fetal poses are allowed during scanning. Gestational age ranges from 19 to 31 weeks, much broader than \cite{li2018standard,namburete2014diagnostic}. Average volume size of our dataset is 270$\times$207$\times$235 and unified voxel size is 0.5$\times$0.5$\times$0.5mm$^{3}$. A sonographer with 5-year experience provided manual annotation of landmarks and standard planes for all the volumes. We then randomly split the dataset into 330 and 100 volumes for training and testing. \par
		
We implemented our framework in \textit{PyTorch}, using a standard PC with a NVIDIA TITAN X(PASCAL) GPU. We trained the DDQN with Adam optimizer (learning rate=5e-5) for 100 epochs(about 4 days). Replay-buffer is set as 15000. Target network copies the parameters of current network every 2000 iterations. For training RNN variants (vanilla RNN and LSTM), optimizer is Adam with L1 regression loss, batch size=100, hidden size=64 and epoch=200 (about 15$mins$). The starting planes for training DDQN are randomly initialized around the ground truth plane within an angle range of $\pm25^{\circ}$ and distance range of $\pm10mm$. The range is deterimined by the average plane localization error of atlas based registration. For landmark detection (Adam optimizer, batch size=1, learning rate=0.001, moment is 0.5, epoch=40), limited by GPU memory, US volume is resized as 0.4 times for training. Gaussian maps of landmarks are generated as ground truth. L2 loss is used for training. Iterative Closest Point algorithm is used for the rigid registration between testing case and atlas. \par

\begin{table}[!htb] \caption {Quantitative evaluation of our proposed framework.} \label{table:quanti_metric}
	\centering
	\scriptsize{
	\begin{tabular}{c|c|c|c|c|c|c}
		\toprule[2pt]		
		\multirow{2}{*}{\bf{Method}} & \multicolumn{3}{c|}{\bf{TC}}  & \multicolumn{3}{|c}{\bf{TT}}\\
		\cline{2-7}
		&Ang($^{\circ}$)\textdownarrow 	&Dis(mm)\textdownarrow 			&SSIM\textuparrow	&Ang($^{\circ}$)\textdownarrow 	&Dis(mm)\textdownarrow 	&SSIM\textuparrow\\
		\hline
		Regress 			&27.04$\pm$8.40		&4.10$\pm$3.81	&0.672$\pm$0.087	&24.27$\pm$17.05	&7.62$\pm$6.00  	 	&0.507$\pm$0.100\\
		AtlasRegist			&14.14$\pm$7.54		&3.40$\pm$2.28 	&0.681$\pm$0.148  	&13.43$\pm$4.63		&2.62$\pm$1.54 		&0.682$\pm$0.138\\
		RegistRegress 		&12.44$\pm$7.78	 	&\textcolor{blue}{2.18$\pm$2.12}		&0.684$\pm$0.157 	&13.87$\pm$11.77 	&2.80$\pm$2.16 		&0.660$\pm$0.141 \\
		\hline
		DDQN-nA 				&31.54$\pm$24.24	&5.12$\pm$3.67		&0.685$\pm$0.131	&30.44$\pm$24.43	&5.03$\pm$3.82    	&0.615$\pm$0.132\\			
		\hline			
		DDQN-maxS 			&11.71$\pm$14.32	&3.53$\pm$2.55	&0.684$\pm$0.165	&12.36$\pm$8.53		&2.95$\pm$2.94			&0.694$\pm$0.154\\
		DDQN-minQ			&10.68$\pm$9.76		&3.40$\pm$2.27	&0.688$\pm$0.165	&10.78$\pm$7.62		&\textcolor{blue}{2.62$\pm$1.54} 			&0.705$\pm$0.163\\
		DDQN-AT(FC) 			&10.36$\pm$9.60	 	&3.40$\pm$2.28	&0.689$\pm$0.165	&9.61$\pm$5.79		&2.66$\pm$1.55 		&0.707$\pm$0.161\\
		DDQN-AT(RNN) 			&9.96$\pm$10.19	 	&3.41$\pm$2.27	&0.691$\pm$0.167	&9.53$\pm$5.74		&2.64$\pm$1.62   	&0.709$\pm$0.164\\
		DDQN-AT(LSTM) 			&\textcolor{blue}{9.61$\pm$8.97} 		&3.40$\pm$2.77	 	&\textcolor{blue}{0.693$\pm$0.168}	&\textcolor{blue}{9.11$\pm$5.56}		&2.66$\pm$2.06 			&\textcolor{blue}{0.709$\pm$0.163}\\
		\toprule[2pt]
	\end{tabular}
}
\vspace{-0.2cm}
\end{table}

\begin{figure}[b]
	\centering
	\includegraphics[width=1.0\textwidth]{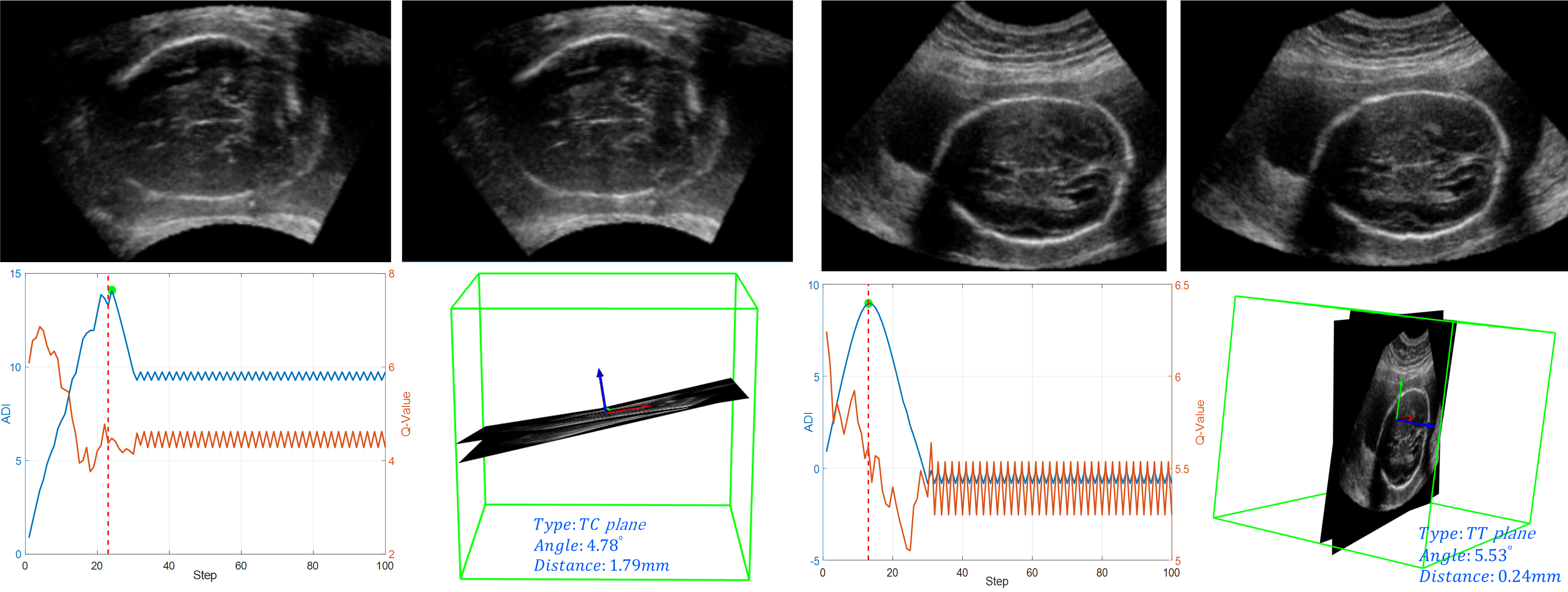}
	\caption{TC (left) and TT (right) results. \textit{Top row:} ground truth (left) and predicted (right) plane. \textit{Bottom row:} left, active termination step (dotted red line) compared to optimal step in green dot, 3D visualization of ground truth and predicted plane (right).}
	\label{fig:visualization}
	\vspace{-0.5cm}
\end{figure}

\subsubsection{Quantitative and Qualitative Analysis:}
The efficacy of our proposed method was validated with 100 US volumes and results were demonstrated in Table \ref{table:quanti_metric}. We adopt both spatial and content similarities to evaluate the performance, including the dihedral angle between two planes (Ang), difference of Euclidean distance to origin (Dis), and Structural Similarity Index (SSIM).
\begin{itemize}
	\item Firstly, the proposed RL agent (DDQN-AT) has good performance on localizing two types of standard planes, and outperforms the regression-based method (Regress), the registration-based method (AtlasRegist) and their combination (RegistRegress). This can be attributed to the active interaction procedure of the agent that searches along the trajectory towards the optimal plane.
	\item Secondly, as can be clearly drawn from the table, when the landmark-aware space alignment module is employed on DDQNs as a warm start, they achieve significantly better performance on standard plane localization than the method without alignment (DDQN-nA). Besides, the proposed space alignment module can also be deployed in the regression model and lead to clear improvement (RegistRegress).
	\item Thirdly, the proposed active termination can lead to better localization with much less inference iterations. Compared to other termination policies, such as maximum step (DDQN-maxS) and minimum Q-value (DDQN-minQ), our AT based methods generally give better localization performances. Among them, DDQN-AT (LSTM) shows the best results, since it has stronger capacity in learning from the Q-value sequence. More importantly, with AT module equipped, our RL-agent requires an average of 13 steps to localize the standard planes, in comparison with 100 steps that no AT module was employed. Given the fact that such iteration steps cost most computation, the AT module will definitely improve the efficiency of the RL agent.
\end{itemize}
	
In Fig. \ref{fig:visualization}, we visualize two testing results of DDQN-AT (LSTM) for TC and TT plane localization. Compared from image content and spatial relationship, for both tasks, our method accurately captures the plane, which is very close to the ground truth. Our active termination strategy also presents the ability to learn from the Q-value sequence and hits the optimal termination step (green dot) for large angle and distance improvement (ADI). \par
	
\section{Conclusion}
We proposed a general framework for standard plane localization in 3D US with a RL agent. We use a landmark-aware alignment model to exploit prior information about the standard planes from the atlas and provide the agent with an effective warm starting point. In addition, we devise a RNN-based active termination strategy to indicate the agent to stop once the optimal plane is localized, therefore improving its accuracy and efficiency. Experiments on our in-house large dataset validate the efficacy of our method and reveal its great potential for future practical applications. \par
	
\subsubsection{Acknowledgments:}
The work in this paper was supported by the grant from National Natural Science Foundation of China (No. 61571304), Shenzhen Peacock Plan (No. KQTD2016053112051497, KQJSCX20180328095606003), Medical Scientific Research Foundation of Guangdong Province, China (No. B2018031) and National Natural Science Foundation of China (Project No. U1813204). \par
	
%
\bibliographystyle{splncs}
\bibliography{refs}

\begin{thebibliography}{10}
\providecommand{\url}[1]{\texttt{#1}}
\providecommand{\urlprefix}{URL }

\bibitem{alansary2018automatic}
Alansary, A., Le~Folgoc, L., et~al.: Automatic view planning with multi-scale
  deep reinforcement learning agents. In: MICCAI. pp. 277--285. Springer (2018)

\bibitem{chykeyuk2013class}
Chykeyuk, K., Yaqub, M., Noble, J.A.: Class-specific regression random forest
  for accurate extraction of standard planes from 3d echocardiography. In:
  International MICCAI Workshop on Medical Computer Vision. pp. 53--62.
  Springer (2013)

\bibitem{ghesu2019multi}
Ghesu, F.C., et~al.: Multi-scale deep reinforcement learning for real-time
  3d-landmark detection in ct scans. IEEE TPAMI  41(1),  176--189 (2019)

\bibitem{li2018standard}
Li, Y., Khanal, B., et~al.: Standard plane detection in 3d fetal ultrasound
  using an iterative transformation network. In: MICCAI. pp. 392--400. Springer
  (2018)

\bibitem{lorenz2018automated}
Lorenz, C., Brosch, T., et~al.: Automated abdominal plane and circumference
  estimation in 3d us for fetal screening. In: Medical Imaging 2018: Image
  Processing. vol. 10574, p. 105740I (2018)

\bibitem{mnih2015human}
Mnih, V., Kavukcuoglu, K., et~al.: Human-level control through deep
  reinforcement learning. Nature  518(7540),  529 (2015)

\bibitem{namburete2014diagnostic}
Namburete, A.I., Stebbing, R.V., Noble, J.A.: Diagnostic plane extraction from
  3d parametric surface of the fetal cranium. In: MIUA. pp. 27--32 (2014)

\bibitem{ronneberger2015u}
Ronneberger, O., Fischer, P., Brox, T.: U-net: Convolutional networks for
  biomedical image segmentation. In: MICCAI. pp. 234--241. Springer (2015)

\bibitem{ryou2016automated}
Ryou, H., Yaqub, M., et~al.: Automated 3d ultrasound biometry planes extraction
  for first trimester fetal assessment. In: MLMI. pp. 196--204. Springer (2016)

\bibitem{schmidt2019offset}
Schmidt-Richberg, A., Schadewaldt, N., et~al.: Offset regression networks for
  view plane estimation in 3d fetal ultrasound. In: Medical Imaging 2019: Image
  Processing. vol. 10949, p. 109493K (2019)

\bibitem{ddqn2016}
Van Hasselt~H, Guez~A, S.D.: Deep reinforcement learning with double
  q-learning. In: AAAI. pp. 1234--1241 (2016)

\end{thebibliography}

\end{document}